\documentclass[preprint,aps,superscriptaddress,nofootinbib]{revtex4-1}
\usepackage{graphicx}
\usepackage{bm}       
\usepackage{amssymb}
\usepackage[toc,page]{appendix}
\usepackage{braket}
\newcommand{\del}{\partial}

\usepackage{amsmath}

\begin{document}

\title{A discrete discontinuity between the two phases of gravity}

\author{Sandipan Sengupta}
\email{sandipan@phy.iitkgp.ac.in}
\affiliation{Department of Physics, Indian Institute of Technology Kharagpur, Kharagpur-721302, INDIA}

\begin{abstract}
%When tetrad (metric) fields are not invertible, the standard canonical formulation of gravity theory cannot be adopted as it is. Here a Hamiltonian analysis of gravity for noninvertible tetrad is presented. We find that this phase exhibits three local degrees of freedom, in constrast with the Einsteinian phase. This unravels a discrete discontinuity in the limit to a vanishing tetrad determinant. For the case of vanishing lapse, fixing the torsional gauge-freedom results in a gravity theory without a Hamiltonian constraint. Any state functional invariant under the internal gauge rotations and spatial diffeomorphisms is a formal solution of the associated quantum theory. The framework here provides a Hamiltonian basis to analyze a physical singularity, which corresponds to a zero of the tetrad determinant in a curved spacetime.
%%%%%%%%
When tetrad (metric) fields are not invertible, the standard canonical formulation of
gravity cannot be adopted as it is. Here we develop a Hamiltonian theory of gravity for
noninvertible tetrad. In contrast to Einstein gravity, this phase is found to exhibit three local degrees of freedom. This reflects a discrete discontinuity in the limit of a vanishing tetrad determinant. For the particular case of vanishing lapse, the Hamiltonian constraint disappears from the classical theory upon fixing the torsional gauge-freedom. Any state functional invariant under the internal gauge rotations and spatial diffeomorphisms is a formal solution of the associated quantum theory. The formulation here provides a Hamiltonian basis to analyze gravity theory around a physical singularity, which corresponds to a zero of the tetrad determinant in curved spacetime.
\end{abstract}
%%%%%%%%%%%%%%%%%%%%%%%%%%%%%%%%%%%%%%%%%%%%%%%%%%%%%%%%%%%

\maketitle
%%%%%%%%%%%%%%%%%%%%%%%%%%%%%%%%%%%%%%%%%%%%%%%%%%%%%%%%%%%%%%%%%%%%%%%

\section{Introduction}
The canonical quantization programme for gravity theory is built upon invertible tetrad (metric) fields. Although the assumption of invertibility might seem natural from the classical perspective, it is rooted in the notion of a smooth geometry which need not truly reflect the spacetime around a singularity. 
%For instance, the analysis of Belinski-Khalatnikov-Lifshitz \cite{bkl} concludes that the spacetime close to the cosmological singularity approaches a metric that is degenerate everywhere. 
In a generic treatment of the quantum spacetime, supposed to be thriving on wild fluctuations, there is no reason to adopt this restrictive assumption anyway. Such a viewpoint has been explored through a number of ideas earlier. A particularly intriguing example is Hanson and Regge's work on torsion foam \cite{regge}, somewhat broader in context than Hawking's spacetime foam \cite{hawking}. Their proposal anticipates a dynamical theory of Euclidean (`quantum') gravity where a vanishing tetrad phase exhibiting torsion vortices should become important, analogous to the normal state in a superconductor dominated by Abrikosov vortices. 

In general, the two possible phases of first-order gravity theory is defined in terms of the tetrad (metric) determinant, which could be either zero or non-zero for a given spacetime solution \cite{tseytlin}. 
%Since this determinant is a four-density, a vanishing determinant is a coordinate invariant feature as long as the coordinate transformations are non-singular. 
These two classes of solutions are mutually exclusive. 

From a classical as well as a quantum perspective, it thus seems important to set up a Hamiltonian form of gravity which could encompass invertible as well as noninvertible tetrad fields. This programme remains incomplete so far. We take up this task here.

To emphasize, we set up a Hamiltonian theory of gravity when the four-determinant of the tetrad vanishes everywhere. The standard ADM parametrization, built upon the inverse tetrad fields, is not a suitable framework in this regard. Although it had been observed earlier that gravity theories could admit a vanishing lapse function as a solution to the Hamiltonian equations of motion \cite{peldan}, the canonical framework associated with such discussions involves invertible tetrad fields to begin with. Such an approach must be avoided in the context of a noninvertible phase. It is in fact essential to first define a convenient set of canonical variables without assuming invertibility of tetrad fields and then obtain the resulting canonical structure of the non-invertible phase, as done here. 

Among the essential details, the associated constraint algebra and the number of local degrees of freedom reveal critical differences with Hilbert-Palatini gravity in the invertible phase.
 Of particular interest is the eventual disappearance of the Hamiltonian (scalar) constraint for vanishing lapse. This finally leads to a first-class system of constraints associated with the internal (gauge)  rotations and spatial diffeomorphisms only.
 
 The potential relevance of degenerate spacetime geometries in gravity theory had been argued over frequently in the literature. An important highlight along these lines is the work by Horowitz \cite{horowitz}, interpreting such spacetimes as possible mediators of topology change. In principle, these geometries should be included in quantum gravity as saddle points of the path integral provided they have finite action. Further studies later on had provided a complete geometric classification of all possible spacetime solutions of the noninvertible phase of first-order gravity, elucidating their intriguing connection with Thurston's model three-geometries \cite{kaul}. Higher dimensional degenerate geometries have appeared in the recent idea of extra dimensions of vanishing proper length, which have remarkably distinctive properties in contrast with the Kaluza-Klein type extra dimensions, and could be relevant in the context of the `dark matter' problem  and also in formulations of well-defined four-dimensional effective Einstein-Gauss-Bonnet (4DEGB) theories \cite{sandipan,sandipan1}.
 
 In some formulations of canonical quantum gravity, geometries with a degenerate spatial metric  emerge naturally as kinematical `vacuum' states. For instance, in the (complex) Sen-Ashtekar \cite{sen,ashtekar} and the (real) Barbero-Immirzi formalism \cite{barbero,immirzi}, one could think of ground states as the ones annihilated by the spatial triad operator \cite{smolin,jacobson1}. Such descriptions, however, leave open the questions addressed here.
 
Further, our work here clearly demonstrates the differences between gravity theories for $\det e_\mu^I =0$ and for the limit $\det e_\mu^I \rightarrow 0$. This point deserves to be emphasized, given that it had often been
missed in the existing literature. Historically, a special case of this limit was first introduced in the metric formulation through a zero signature condition \cite{teitelboim,henneaux} imposed on the original ADM constraints. This leads to what is now christened as the `electric' Carrollian limit of gravity in the literature \cite{hartong,bergshoeff,henneaux1}, restricted to the particular context of metric gravity. 
% This issue is particularly relevant in the context of the Carrollian (Levy Leblond-Sen Gupta) limits of gravity. The fact that degenerate metrics play an important role in this context was illustrated by Henneaux \cite{henneaux} in his formulation of the `electric' Carroll limit within the metric (ADM) formalism of Hamiltonian gravity. 
Recently, it has been demonstrated how the so-called `electric' and `magnetic' Carrollian limits could emerge from
the same Hilbert-Palatini action, and a detailed Hamiltonian analysis within the first-order framework was also set up \cite{sengupta}. 
%Through a constraint analysis,  this formulation was shown to be associated with two local degrees of freedom. 
%This is the same number which is obtained in the case of Henneaux's `electric' limit in metric gravity \cite{henneaux}, even though the respective Hamiltonian structures exhibit nontrivial differences.  
%These studies show that the degrees of freedom count in these singular limits are the same as in original Hilbert-Palatini gravity.

In the next section, we proceed with the analysis of the Hamiltonian theory in question. The full set of constraints, the Poisson algebra and the solutions are presented first for the case when the zero eigenvalue lies along a timelike direction (vanishing lapse). 
%The degeneracy of the tetrad is implemented as one of the Hamiltonian equations of motion. 
Next, we briefly analyze the case where the null eigenvalue resides in a spatial direction, leading to a degenerate triad (nonvanishing lapse). 
%This is followed by a demonstration of the Hamiltonian equations of motion here to the Lagrangian ones. 
The concluding section contains a summary and some perspective.

\section{Hamiltonian theory in the vanishing determinant phase}
In the first order form based on an $SO(3,1)$ internal symmetry group, pure gravity could be described in terms of the tetrad  and spin-connection fields $e_\mu^I$ and $\omega_\mu^{~IJ}$, respectively. The associated action does not depend on the inverse tetrad (metric).
%, and the resulting field equations admit spacetime solutions exhibiting both phases with $\det e_\mu^I=0$ and $\det e_\mu^I\neq 0$: 
In order to  formulate a Hamiltonian theory of (vacuum) gravity at the phase $\det e_\mu^I=0$, we begin with the Hilbert-Palatini Lagrangian density: 
\begin{equation} \label{L0}
{\cal L}(e,\omega) ~ = ~ \frac{1}{8\kappa}\epsilon^{\mu\nu\alpha\beta}\epsilon_{IJKL}e_\mu^I e_\nu^J R_{\alpha\beta}^{~ KL}(\omega) 
\end{equation}
Here $\kappa$ is the gravitational coupling (to be set to unity in the following), $\epsilon^{\mu\nu\alpha\beta}$ 
%($\epsilon^{txyz}=1,~\epsilon_{txyz}=\det g_{\mu\nu}$) 
is the completely antisymmetric symbol with density weight one and $R_{\mu\nu}^{~~~ IJ}=\del_{[\mu}\omega_{\nu]}^{~IJ}+\omega_{[\mu}^{~IK}\omega_{\nu]K}^{~~~J}$ is the $SO(3,1)$ field strength. The internal space is associated with a Minkowskian metric ($-+++$) and an antisymmetric symbol $\epsilon_{IJKL}$.

We enforce a space-time split of the above, leading to ($\epsilon^{tabc}\equiv \epsilon^{abc}$):
\begin{eqnarray}\label{L1}
{\cal L}= \frac{1}{4}\epsilon^{abc}\epsilon_{IJKL}e_b^K e_c^L\del_t\omega_a^{~IJ}+\frac{1}{4} e_t^I\left[ \epsilon^{abc}\epsilon_{IJKL}e_a^J R_{bc}^{~KL}\right]+\frac{1}{4}\omega_t^{~IJ}D_{a}\left[\epsilon^{abc}\epsilon_{IJKL}e_b^K e_c^L\right]
\end{eqnarray}
The momenta $P^a_I$ and $\Pi^a_{~IJ}$ (conjugate to $e_a^I$ and $\omega_a^{~IJ}$, respectively), which could be read off from the above, are subject to the following primary constraints:
\begin{eqnarray}\label{pc}
P^a_I\approx 0,~\chi^a_{~IJ}~\equiv~\Pi^{a}_{~IJ}-\frac{1}{2}\epsilon^{abc}\epsilon_{IJKL}e_b^K e_c^L\approx 0.
\end{eqnarray}
Further, the multiplier fields $e_t^I$ and $\omega_t^{~IJ}$ in eq.(\ref{L1}) correspond to vanishing momenta, respectively:  \begin{eqnarray}\label{trivial}
P_I\approx 0,~\Pi_{IJ}\approx 0.
\end{eqnarray}
The latter in turn imply the following secondary constraints:
\begin{eqnarray}\label{D}
&&C_I\equiv -\frac{1}{4}\epsilon^{abc}\epsilon_{IJKL}e_a^J R_{bc}^{~KL}\approx 0,~G_{IJ}\equiv -\frac{1}{2}D_{a}\left[\epsilon^{abc}\epsilon_{IJKL}e_b^K e_c^L\right]\approx 0.
\end{eqnarray}
Thus, we are led to the following primary Hamiltonian density:
\begin{eqnarray}\label{H}
{\cal H}=e_t^I C_I+\frac{1}{2}\omega_t^{~IJ}G_{IJ}+\mu_a^{I}P^a_{I}+\mu_a^{~IJ}\chi^a_{~IJ},
\end{eqnarray}
where $\mu_a^I$ and $\mu_a^{~IJ}$ are the Lagrange multipliers.
%Note that the space-time decomposition above applies to invertible as well as non-invertible tetrad fields. In the first case, it is possible to invert the last constraint in (\ref{pc}) to eliminate the fields $e_a^I$ completely in terms of the momenta $\Pi^a_{~IJ}$. In the latter case, however, this is not possible and one requires a different approach.

\subsection{Constraint algebra}

Given the basic Poisson brackets:
\begin{eqnarray}
[e_a^I(x),P^b_J(y)]=\delta_a^b \delta^3 (x-y),~[\omega_a^{~IJ}(x),\Pi^b_{~KL}(y)]=\delta_a^b \delta^{[I}_K\delta^{J]}_L \delta^3(x-y),
\end{eqnarray}
the Poisson algebra between the constraints $(C_I,G_{IJ},P^a_I,\chi^a_{~IJ})$ is obtained as below:
\begin{eqnarray}\label{alg}
&&\left[\int \Lambda^I C_I,\int \Omega^J C_J\right]=0,\nonumber\\
&&\left[\int  \Lambda^I C_I,\int \Omega^{KL}G_{KL}\right]=0,\nonumber\\
&&\left[\int \Lambda^I C_I,\int \Omega_a^{JK} \chi^a_{~JK}\right]=-\int \epsilon^{abc}\epsilon_{IJKL}\Lambda^K e_c^L D_a\Omega_b^{IJ},\nonumber\\
&&\left[\int \Lambda^I C_I,\int \Omega_a^{K} P^a_{K}\right]=\frac{1}{4}\int \epsilon^{abc}\epsilon_{IJKL}\Omega_a^{I}\Lambda^J R_{bc}^{~KL},\nonumber\\
%&&\left[\int \Lambda^I C_I,\int \Omega^a {\cal E}_a\right]=0,\nonumber\\
%&&\left[\int \Lambda^I C_I,\int \Omega \Pi_\phi\right]=-\frac{1}{3}\int \epsilon^{abc}\epsilon_{IJKL}\Omega \Lambda^I e_a^J e_b^K e_c^L,\nonumber\\
&&\left[\int \Lambda^{IJ}G_{IJ},\int\Omega^{KL}G_{KL}\right]=0,\nonumber\\
&&\left[\int \Lambda^{IJ}G_{IJ},\int\Omega_a^{KL} \chi^a_{KL}\right]=  2\int \epsilon^{abc}\epsilon_{IJKL}\Lambda_{KM} \Omega_a^{IJ}e_b^M e_c^L ,\nonumber\\
%&&\left[\int\Lambda^{IJ}G_{IJ},\int \Omega^a {\cal E}_a\right]= 0,    \nonumber\\
&&\left[\int\Lambda^{IJ}G_{IJ},\int \Omega_a^{K} P^a_{K}\right]=   \int \epsilon^{abc}\epsilon_{IJKL}e_b^K  \Omega_c^{L}D_a\Lambda^{IJ}, \nonumber\\
&&\left[\int \Lambda_a^{IJ} \chi^a_{~IJ},\int \Omega_b^{KL} \chi^b_{~KL}\right]=0, \nonumber\\
&&\left[\int \Lambda_a^{IJ} \chi^a_{IJ},\int\Omega_b^{L} P^b_{L}\right]= \int \epsilon^{abc}\epsilon_{IJKL}\Lambda _a^{KL}\Omega_b^{J} e_c^I,    \nonumber\\
%&&\left[\int \Lambda_a^{IJ} \chi^a_{IJ},\int \Omega^b {\cal E}_b\right]=0, \nonumber\\
&&\left[\int \Lambda_a^{I} P^a_{I},\int \Lambda_b^{J} P^b_{J}\right]=0. \nonumber\\
%&&\left[\int \Lambda_a^{I} P^a_{I},\int \Omega^b {\cal E}_b\right]=-\int \epsilon^{bcd}\epsilon_{IJKL}\Omega^a\left[\Lambda_a^I e_b^J-3\Lambda_b^I e_a^J\right]e_c^K e_d^L,\nonumber\\
%&&\left[\int \Lambda^a {\cal E}_a ,\int \Omega^b {\cal E}_b\right]=0.\nonumber\\
 %&&\left[\int \Lambda {\cal D},\int \Omega_a^{K} P^a_{K}\right]=3\int \epsilon^{abc}\epsilon_{IJKL}\Lambda e_t^I \Omega_a^{J}e_b^K e_c^L.
\end{eqnarray}
In the above, $(\Lambda,\Omega)$ are arbitrary smearing functions. Note that the algebra reflects the existence of several second-class pairs.

%Even though not essential, let us also display the brackets involving $C^{ab}$ as introduced in (\ref{C-ab}):
%\begin{eqnarray}\label{alg2}
%&&\left[\int \Lambda^I C_I,\int \Omega_{ab}C^{ab}\right]= 4\int\epsilon^{abc}\Omega_{ad}\left[\Lambda^I\Pi^d_{~IJ} D_be_c^J+e_c^J D_b \Lambda^I \chi^d_{~IJ}\right],\nonumber\\
%&&\left[\int\Lambda^{IJ}G_{IJ},\int\Omega_{ab}C^{ab}\right]=4\int\epsilon^{IJML}\Omega_{ad}\Lambda_{KM}\Pi^a_{~KL} \chi^d_{~IJ},\nonumber\\
%&&\left[\int\Lambda_a^{I} P^a_{I},\int\Omega_{bc}C^{bc}\right]=0,\nonumber\\
%&&\left[\int\Lambda_{ab}C^{ab},\int\Omega_{cd}C^{cd}\right]=0.
%\end{eqnarray}

\subsection{Vanishing tetrad determinant as a Hamiltonian solution}
From inspection of the constraints $\chi^a_{~IJ}\approx 0$, we introduce the following variables in order to obtain a solution \cite{sa,kaul1}:
\begin{eqnarray}\label{E}
\Pi^a_{~0i}=E^a_i,~\Pi^a_{~ij}=\chi_{[i}E^a_{j]},
\end{eqnarray}
where:
\begin{eqnarray}\label{E1}
e_a^0=-e_a^i \chi_i
\end{eqnarray}
The first equation in (\ref{E}) defines nine solutions for the momenta $\Pi^a_{~0i}$ through the fields $E^a_i$. The second one eliminates the six redundant components of $\Pi^a_{~ij}$, as the RHS of this equation has only three new fields $\chi^i$. Eqn.(\ref{E1}) expresses the fact that the three coordinates $e_a^0$ are not independent. Altogether, these constitute the eighteen solutions to the eighteen constraints $\chi^a_{~IJ}$.

%\subsection{Symplectic form}
Using these fields, we may rewrite symplectic form as:
\begin{eqnarray*}
\Omega=\Pi^a_{~0i}\del_t \omega_a^{~0i}+\frac{1}{2}\Pi^a_{~ij}\del_t \omega_a^{~ij}+P^a_I \del_t e_a^I=E^a_i\del_t Q_a^i+\zeta^i \del_t \chi_i+\hat{P}^a_i \del_t e_a^i
\end{eqnarray*}
where $Q_a^i\equiv \omega_a^{~0i}-\chi_j \omega_a^{~ij},~\zeta^i\equiv -E^a_j \omega_a^{~ij}-P^a_0 e_a^i,~\hat{P}^a_i\equiv P^a_i-P^a_0\chi_i$. 
%Thus, starting with the thirty canonical pairs, we end up with twenty-one pairs. 
Implementing the constraints $P^a_I=0$ strongly alongside, we obtain:
\begin{eqnarray}\label{Omega}
\Omega=E^a_i\del_t Q_a^i+\zeta^i \del_t \chi_i
\end{eqnarray}
Note that the fields $E^a_i$ are in fact the densitized triad:
\begin{eqnarray}\label{e}
\hat{e}^a_i=\sqrt{E} E^a_i,
\end{eqnarray}
where the inverse of the triad are defined as $\hat{e}^a_i$ ($\frac{1}{E}\equiv \det E^a_i=e^2$) \footnote{To emphasize, these in general are not the same as  the spatial components of tetrad inverse $e^a_i$, which exist only for invertible tetrad fields.}. 

%\subsection{Nonvanishing torsion in vacuum}

Clearly, from (\ref{Omega}), the only dynamical components of the fields $\omega_a^{~ij}$ are $\zeta^i$. The rest could be parametrized as a field $\bar{N}^{kl}=\bar{N}^{lk}$ \cite{sa,kaul1}: 
\begin{equation*}
\omega_a^{~ij}=\frac{1}{2}E_{a}^{[i}\zeta^{j]}+\epsilon^{ijk}E_a^l \bar{N}^{kl}
\end{equation*}
Since the coordinates $\bar{N}_{kl}$ are absent in the symplectic form, their conjugate momenta must vanish: ($\bar{\pi}_{kl}\approx 0$). Further, these must be preserved in time, implying:
\begin{eqnarray}\label{torsion}
[H,\bar{\pi}_{mn}]\approx 0\approx \epsilon^{ebc}\left[e_t^0 e_c^m D_a e_b^n-2e_t^i e_c^m(\delta_i^n D_a e_b^0+\chi_l \delta_i^{[n} D_a e_b^{l]})\right]+[m\leftrightarrow n]
\end{eqnarray}
Let us consider the following secondary constraints emerging from the above:
\begin{eqnarray}\label{dege}
e_t^0\approx 0,~e_t^i \approx 0.
\end{eqnarray}
This class of solutions correspond to the noninvertible (non-Einsteinian) phase of first order gravity.
Note that for this case, the connection components $\bar{N}^{kl}$ are left undetermined, reflecting that torsion in vacuum is nonvanishing unlike in Einstein gravity. 
The solution to the spatial connection may as well be rewritten as:
\begin{eqnarray}\label{N}
\omega_a^{~ij}=\bar{\omega}_a^{~ij}(E)+\epsilon^{ijk}E_a^l N^{kl}
\end{eqnarray}
where the new fields $N^{kl}$ differ from the older ones by a symmetric piece. This indeterminacy of the connection fields brings in important consequences for the Hamiltonian constraint in particular, to be discussed later. 

For $e_t^I\neq 0$, on the other hand, eq.(\ref{torsion}) leads to a different set of secondary constraints which correspond to invertible tetrad. The fields $\bar{N}^{kl}$ in that case are completely determined as:
$\bar{N}^{kl}=\frac{1}{2}\epsilon^{ijk}E_l^a \left[\bar{\omega}_a^{~ij}(E)-E_a^{i}\zeta^{j}\right]$,  where $\bar{\omega}_a^{~ij}(E)$ are the torsionless components of the spatial connection. This is equivalent to the vanishing of torsion ($D_{[a} e_{b]}^i=0=D_{[a} e_{b]}^0$) in vacuum gravity \cite{peldan,sa}. We would not elaborate any further on this well-studied case here.

The resulting expressions for the set of constraints in the noninvertible tetrad phase are displayed below:
\begin{eqnarray}
&& P_I\approx 0,~\Pi_{IJ}\approx 0,~\pi_{kl}\approx 0,\nonumber\\
&& G^i_{boost}\equiv G^{0i}=\del_aE^a_i+ \chi_{[i}E^a_{k]}Q_a^k-(\delta^i_{l}+\chi^i \chi_l)\zeta^l\approx 0,\nonumber\\
&& G^i_{rot}\equiv\frac{1}{2}\epsilon^{ijk}G_{jk}=\epsilon^{ijk}[\del_a (\chi_j E^a_k)+\chi^k \zeta^j- E^a_{k}Q_a^j)]\approx 0,\nonumber\\
&& C_0=-\frac{\sqrt{E}}{2}E^{[a}_i E^{b]}_j\left[\del_a\omega _b^{~ij}+\omega_a^{~ik}\omega_b^{~kj}+Q_a^i Q_b^j+2\chi_k Q_b^{j}\omega_a^{~ik}+\chi_k\chi_l\omega_{a}^{~ik}\omega_{b}^{~jl}\right]\approx 0,\nonumber\\
%%%%
&& C_i=\sqrt{E}E^{[a}_i E^{b]}_j\left[\del_a Q_b^{k}+\omega_a^{~kl}Q_b^l+\del_a(\chi_l \omega_b^{~kl})
+\chi_l\omega_{a}^{~ml}\omega_{b}^{~mk}\right]\nonumber\\
&&- \frac{\sqrt{E}}{2}\left(\chi_i E^{[a}_j E^{b]}_k+\chi_j E^{[a}_k E^{b]}_i+\chi_k E^{[a}_i E^{b]}_j\right)[\del_a\omega _b^{~jk}+\omega_a^{~jm}\omega_b^{~mk}+Q_a^j Q_b^k+2\chi_m Q_b^{k}\omega_a^{~jm}\nonumber\\
&&+\chi_m\chi_n\omega_{a}^{~jm}\omega_{b}^{~kn}]\approx 0,\nonumber\\
&& e_t^I\approx 0, 
\end{eqnarray}
where, in the first line involving only the (surviving) primary constraints we have defined $\pi_{kl}$ as the conjugate momenta of $N^{kl}$ introduced in (\ref{N}). 

The above set of constraints define the full Hamiltonian theory of gravity at the $\det e_\mu^I=0$ phase.

\subsection{Time-gauge constraints}
It is possible to fix the boost freedom here through the time gauge: $\chi_i=0$ \cite{holst}. Using the corresponding solution of the boost constraint given by $\zeta_i=\del_a E^a_i$, the constraints then attain a much simpler form:
\begin{eqnarray}\label{C}
&& P_I\approx 0,~\Pi_{IJ}\approx 0,~\pi_{kl}\approx 0,\nonumber\\
&& G^i_{rot}=-\epsilon^{ijk} Q_a^j E^a_{k}\approx 0,\nonumber\\
&& C_0=-\frac{\sqrt{E}}{2}\left[E^{[a}_i E^{b]}_j(\bar{R}_{ab}^{~ij}(\bar{\omega})+Q_a^i Q_b^j)+N^{ij}N_{ji}-N^i_{~i} N^j_{~j}\right]\approx 0,\nonumber\\
&& C_i=\sqrt{E}\left[E^{[a}_i E^{b]}_k\bar{D}_a Q_b^k-N_{ki}G^k_{rot}\right]\approx 0,\nonumber\\
&& e_t^I\approx 0,
\end{eqnarray}
where we have defined $\bar{R}_{ab}^{~ij}(\bar{\omega}(E))=\del_{[a}\bar{\omega}_{b]}^{~ij}+\bar{\omega}_{[a}^{~ik}\bar{\omega}_{b]}^{~kj}$ as the field-strength and $\bar{D}_a$ as the covariant derivative with respect to the torsionless (spatial) connection components $\bar{\omega}_a^{~ij}(e)$. In comparison with the time-gauge constraints of the invertible phase \cite{peldan}, we note that the vector and rotation constraints exhibit exactly the same canonical form (in a weak sense). The Hamiltonian (scalar) constraint however gets modified.\footnote{Note that in time-gauge, $C^0$ and $C^i$ are related to the Hamiltonian and diffeomorphism constraints respectively as: $H=C_0,~H_a=\frac{1}{\sqrt{E}}E_a^i C_i$.}

%Owing to the $N^{ij}$-dependent terms, $C^0$ forms a second-class pair with $G^{i}_{rot}$ in time gauge. Along with $C^i$, this leads to one second-class pair and five first-class constraints among nine canonical pairs $(Q_a^i,E^b_j)$. This reconfirms the fact that the degrees of freedom per spacetime point is three, exactly the same as found earlier without fixing any gauge.

\subsection{Degrees of freedom: A discrete discontinuity in gravity}

The set $(C^0,\pi^{ij})$ contains a second-class pair along with five first-class constraints. In addition, there are four second-class pair $(e_t^I,P_J)$ and twelve first-class constraints in $(\Pi^{IJ},G^i_{rot},C^j)$. Since the final phase space is defined by the canonical pairs $(e_t^I,P_J)$, $(\omega_t^{~IJ},\Pi^{KL})$, $(Q_a^i,E^b_j)$ and $(N_{kl},\pi^{mn})$, this leads to $25-(5+17)=3$ local degrees of freedom.

This result for the vanishing tetrad determinant reveals yet another important contrast to the invertible phase of first-order gravity in vacuum. The latter is known to exhibit two local degrees of freedom. In the limiting case of $\det e_\mu^I\rightarrow 0$, which is different from the case of an exact degeneracy of tetrad, one obtains two physical polarizations as well, as has been demonstrated recently  in the context of a first order formulation of Carrollian gravity \cite{sengupta}. 

Hence, we conclude that the limit of a vanishing tetrad (metric) determinant in gravity theory exhibits a discrete discontinuity in the number of local degrees of freedom. 

\subsection{Disappearance of the Hamiltonian constraint}
It is imperative to solve the second-class constraints before proceeding to a canonical quantization. By inspection of eq.(\ref{C}), this could be achieved by solving the Hamiltonian constraint to fix the nondynamical scalar component $\xi=(N^{ij}N_{ji}-N^i_{~i} N^j_{~j})$ and setting the associated momenta to zero (strongly):
\begin{eqnarray}\label{car1}
\xi=-E^a_{[i} E^b_{j]}[\bar{R}_{ab}^{~ij}(\bar{\omega}(E))+Q_a^i Q_b^j].
\end{eqnarray}
This implies that the solutions admit any arbitrary scalar curvature for the spatial three-geometry.
The rest of the components of the (redundant) momenta $\pi_{kl}$ may as well be set to zero, which leaves the remaining constraints unaffected. 

To summarize, the canonical theory finally is defined by the pair $(Q_a^i,E^b_j)$, subject to the first-class constraints $(G^i_{rot},C^j)$, but having no scalar (Hamiltonian) constraint.

Before concluding this section, let us note that inclusion of a cosmological constant in this analysis is straightforward. It simply shows up as an additive constant in the arbitrary field $\xi$, and could be absorbed away into a redefinition without affecting any of the physical results. 
In other words, the effective cosmological constant in this phase of gravity is zero.

\section{Hamiltonian gravity with a degenerate spatial triad}

For completeness, we now briefly discuss the (only) other possibility associated with a degenerate tetrad, where the null eigenvalue of the tetrad lies along a spatial direction. This implies that the spatial triad is degenerate ($\det e_a^i=0$) while the lapse is nonvanishing. 

In the last section, the degeneracy of tetrad was manifestly obtained as a solution to the Hamiltonian equations of motion. Equivalently, such a feature could also be enforced directly through a Lagrange multiplier field in the Lagrangian density. For convenience, we adopt the latter approach in this section:
 \begin{equation} \label{L0}
{\cal L}(e,\omega,\lambda) ~ = ~ \frac{1}{8}\epsilon^{\mu\nu\alpha\beta}\epsilon^{IJKL}e_\mu^I e_\nu^J R_{\alpha\beta}^{~ KL}(\omega) +\frac{\phi}{12}\epsilon^{\mu\nu\alpha\beta}\epsilon_{IJKL}e_\mu^I e_\nu^J e_\alpha^K e_\beta^L,
\end{equation}
$\phi$ being a multiplier field. The primary constraints read:
\begin{eqnarray}\label{pconstr}
&&P_I\equiv \frac{\del{\cal L}}{\del\dot{e}_t^I}\approx 0,~P^a_I\equiv \frac{\del{\cal L}}{\del\dot{e}_a^I}\approx 0,~\Pi_{IJ}\equiv \frac{\del{\cal L}}{\del\dot{\omega}_t^{IJ}}\approx 0,~\Pi_\phi\equiv \frac{\del{\cal L}}{\del\dot{\lambda}}\approx 0,\nonumber\\
&&\chi^a_{~IJ}\equiv\Pi^a_{~IJ}-\frac{1}{2}\epsilon^{abc}\epsilon_{IJKL}e_b^k e_c^L\approx 0,
\end{eqnarray}
 where $\Pi^a_{~IJ}$ are the momenta conjugate to $\omega_a^{~IJ}$ as earlier. These in turn lead to the following set of secondary constraints:
 \begin{eqnarray}
&&\bar{C}_I\equiv -\epsilon^{abc}\epsilon_{IJKL}e_a^J\left[\frac{1}{4} R_{bc}^{~KL}+\frac{\phi}{3} e_b^K e_c^L\right]\approx 0,\nonumber\\
&&G_{IJ}\equiv -\frac{1}{2}D_{a}\left[\epsilon^{abc}\epsilon_{IJKL}e_b^K e_c^L\right]\approx 0,\nonumber\\
&&{\cal D}=\epsilon^{abc}\epsilon_{IJKL}e_t^I e_a^J e_b^K e_c^L\approx 0.
\end{eqnarray}

The case of a vanishing lapse ($e_t^0= 0$) has already been considered earlier.
For a nonvanishing lapse ($e_t^0\neq 0$), the solution for the degeneracy constraint ${\cal D}\approx 0$ is given by $\epsilon^{abc}\epsilon_{ijk}e_a^i e_b^j e_c^k\approx 0$, reflecting a degenerate triad. This is what concerns us in this section.

The associated constraint algebra is given below:
\begin{eqnarray}\label{alg3}
&&\left[\int \Lambda^I \bar{C}_I,\int \Omega^J \bar{C}_J\right]=0,\nonumber\\
&&\left[\int  \Lambda^I \bar{C}_I,\int \Omega^{KL}G_{KL}\right]=0,\nonumber\\
&&\left[\int \Lambda^I \bar{C}_I,\int \Omega_a^{JK} \chi^a_{~JK}\right]=-\int \epsilon^{abc}\epsilon_{IJKL}\Lambda^K e_c^L D_a\Omega_b^{IJ},\nonumber\\
&&\left[\int \Lambda^I \bar{C}_I,\int \Omega_a^{K} P^a_{K}\right]=\frac{1}{4}\int \epsilon^{abc}\epsilon_{IJKL}\Omega_a^{I}\Lambda^J \left[R_{bc}^{~KL}+4\phi e_b^K e_c^L\right],\nonumber\\
&&\left[\int \Lambda^I \bar{C}_I,\int \Omega {\cal D}\right]=0,\nonumber\\
&&\left[\int \Lambda^{IJ}G_{IJ},\int\Omega^{KL}G_{KL}\right]=0,\nonumber\\
&&\left[\int \Lambda^{IJ}G_{IJ},\int\Omega_a^{KL} \chi^a_{KL}\right]=  2\int \epsilon^{abc}\epsilon_{IJKL}\Lambda_{KM} \Omega_a^{IJ}e_b^M e_c^L ,\nonumber\\
&&\left[\int\Lambda^{IJ}G_{IJ},\int \Omega {\cal D}\right]= 0,    \nonumber\\
&&\left[\int\Lambda^{IJ}G_{IJ},\int \Omega_a^{K} P^a_{K}\right]=   \int \epsilon^{abc}\epsilon_{IJKL}e_b^K  \Omega_c^{L}D_a\Lambda^{IJ}, \nonumber\\
&&\left[\int \Lambda_a^{IJ} \chi^a_{~IJ},\int \Omega_b^{KL} \chi^b_{~KL}\right]=0, \nonumber\\
&&\left[\int \Lambda_a^{IJ} \chi^a_{IJ},\int\Omega_b^{L} P^b_{L}\right]= \int \epsilon^{abc}\epsilon_{IJKL}\Lambda _a^{KL}\Omega_b^{J} e_c^I,    \nonumber\\
&&\left[\int \Lambda_a^{IJ} \chi^a_{IJ},\int \Omega {\cal D}\right]=0, \nonumber\\
&&\left[\int \Lambda_a^{I} P^a_{I},\int \Lambda_b^{J} P^b_{J}\right]=0, \nonumber\\
&&\left[\int \Lambda_a^{I} P^a_{I},\int \Omega {\cal D}\right]=\int \epsilon^{abc}\epsilon_{IJKL}\Omega\Lambda_a^I e_t^J e_b^K e_c^L\nonumber\\
&&\left[\int \Lambda {\cal D} ,\int \Omega {\cal D}\right]=0.
\end{eqnarray}

It is useful to note that from the set of eighteen constraints $\chi^a_{~IJ}\approx 0$, it is possible to isolate a set of six which involve the momenta only ($\chi^a_{~IJ}[18]\equiv (\hat{\chi}^a_{~IJ}[12],~C^{cd}[6])$):
\begin{eqnarray}\label{C-ab}
C^{ab}\equiv \frac{1}{2}\epsilon^{IJKL}\Pi^a_{~IJ}\Pi^b_{~KL}\approx 0.
\end{eqnarray}
This subclass $C^{ab}$ exhibits the following brackets with the rest:
\begin{eqnarray}\label{alg4}
&&\left[\int \Lambda^I \bar{C}_I,\int \Omega_{ab}C^{ab}\right]= 4\int\epsilon^{abc}\Omega_{ad}\left[\Lambda^I\Pi^d_{~IJ} D_be_c^J+e_c^J D_b \Lambda^I \chi^d_{~IJ}\right],\nonumber\\
&&\left[\int\Lambda^{IJ}G_{IJ},\int\Omega_{ab}C^{ab}\right]=4\int\epsilon^{IJML}\Omega_{ad}\Lambda_{KM}\Pi^a_{~KL} \chi^d_{~IJ},\nonumber\\
&&\left[\int\Lambda_a^{I} P^a_{I},\int\Omega_{bc}C^{bc}\right]=0,\nonumber\\
&&\left[\int\Lambda{\cal D},\int\Omega_{bc}C^{bc}\right]=0,\nonumber\\
&&\left[\int\Lambda_{ab}C^{ab},\int\Omega_{cd}C^{cd}\right]=0.
\end{eqnarray}
Let us now consider the time evolution of the primary constraints $C^{ab}$:
\begin{eqnarray}
[H,C^{ab}]=\epsilon^{cd(a}\Pi^{b)}_{~IJ}e_t^I D_c e_d^J\approx 0.
\end{eqnarray}
For a nonvanishing lapse ($e_t^0\neq 0$), this consistency condition is solved by:
\begin{eqnarray}
D^{ab}\equiv\epsilon^{cd(a}\Pi^{b)}_{~0j} D_c e_d^j\approx 0,~e_t^i\approx 0.
\end{eqnarray}
These are additional secondary constraints, beyond which there are no further ones.

The nonvanishing Poisson brackets $[\hat{\chi}^a_{~IJ},P^b_
k],~[C^{ab},D^{cd}], ~[e_t^i,P_j],~[{\cal D},P_0]$ reflect $12+6+3+1=22$ second-class pairs. Further, the set $[\Pi_\phi,\bar{C}_i]$ contain one second-class pair and two first-class constraints. The rest, namely $(\Pi^{IJ},~G_{IJ},~\bar{C}_0)$, represent $6+6+1=13$ first-class constraints. These many constraints among the $18+6+12+4+1=41$ canonical pairs $(\omega_a^{~IJ},\Pi^b_{~KL}),~(\omega_t^{~IJ},\Pi_{KL}),~(e_a^I,P^b_J),~(e_t^I,P_J)$ and $(\phi,\Pi_\phi)$ leaves $41-(23+15)=3$ degrees of freedom per spacetime point.

Thus, we conclude that gravity theory for a degenerate tetrad (metric) exhibits three states of polarizations irrespective of whether the null eigendirection is timelike or spacelike. 

 \section{Conclusions}

Classical gravity theory is known to exhibit spacetime solutions with noninvertible tetrads. However, there exists no Hamiltonian formulation of gravity equipped to address such a phase. This phase and the associated canonical theory is expected to be relevant close to curvature singularities and at short
 distances, where a smooth spacetime metric might not exist. With an aim to systematically investigate the potential importance of this phase, we set up a detailed Hamiltonian theory of gravity for noninvertible tetrad fields.

%We find that the limit to a vanishing tetrad (metric) determinant in gravity theory is discretely discontinuous. 
We find that gravity theory in the $\det e_\mu^I=0$ phase is associated with three local degrees of freedom, irrespective of whether the zero eigenvalue lies along a timelike (vanishing lapse but nondegenerate triad) or spacelike (degenerate triad but nonvanishing lapse) direction. Hence, the limit $\det e_\mu^I\rightarrow 0$ reflects a discrete discontinuity, since for any arbitrarily small but nonvanishing determinant of tetrad the number of physical polarization states remains two. This phenomenon is somewhat analogous to the celebrated vDVZ discontinuity \cite{vdv,z}, where an exactly vanishing mass of the graviton corresponds to five polarization states as opposed to two in the zero mass limit. The analogy, however, is incomplete in the sense that unlike graviton mass, the tetrad determinant is a spacetime field.

The canonical structure of the non-invertible phase differs in a number of important aspects when compared to the (Einsteinian) phase of invertible tetrad gravity. For instance, for a vanishing lapse the Hamiltonian constraint disappears from the theory once the torsional gauge freedom is fixed. This is an inviting feature from the perspective of a canonical quantization. Formally, any functional invariant under internal (gauge) rotations and spatial diffeomorphisms are solutions of the associated quantum constraints (upto regularization and ordering ambiguities). Candidates for such physical states are well-known in the context of metric, connection or loop representation \cite{smolin,smolin1,baez}. An worthwhile question to ask is, whether an Einsteinian metric phase of gravity could be obtained as a perturbation over the quantum solutions of the vanishing determinant phase. 
%,We note that  given the natural abundance of Dirac observables here. 

We note that the geometric field theory of Husain and Kuchar \cite{husain}, introduced as 
a toy model of gravity, has no Hamiltonian constraint to begin with. A worthwhile
question to ask is, whether these field theories could be connected to the more general
formulation here.

%The relatively simpler final form of the canonical theory associated with zero-determinant gravity seems inviting enough for one to delve into the interpretational aspects of the candidate quantum solutions (states). 
%Early universe seems to be an arena where possible effects of the quantum seeds could be relevant. The reasons behind such an expectation are manifold, given the nonstandard causal properties and the natural emergence of a `torsion foam' in this phase. We also expect our analysis to play a role towards a deeper understanding of the cosmological singularity, where the metric of the time-reversed `Universe' appears to approach degeneracy. %The lack of a Hamiltonian theory of gravity based on degenerate tetrad fields had been a serious impediment towards a significant progress along these lines.  

We also find that the cosmological constant, if included in the action for a vanishing determinant, could be absorbed away through a redefinition of the multiplier fields. One wonders whether the absence of an effective cosmological constant in this phase could be connected to its observed smallness in a perturbed (physical) state around it. 

Contexts such as above would help us know if the noninvertible phase of gravity could indeed be ascribed a physical meaning, a question that remains very much open.
 
%Since torsion involves its own length scale, the indeterminacy in connection could have nontrivial imports for perturbative gravity as well.
 
%Further, we have shown that the scalar constraint (for temporal degeneracy) reproduces the `electric' and `magnetic' Carrollian limits of the Hamiltonian constraint as special cases. This reflects an interesting connection between the exact and limiting cases of a vanishing metric determinant, despite the mutual inequivalence of the associated gravity theories owing to the essential discontinuity of  the limit.                 
 
%Our final remark concerns the quantum geometries which could be built upon this analysis. In particular, one is now allowed to consider a quantum vacuum corresponding to a degenerate (or, in an even stronger sense, a trivial) four-metric. As a possible application, it would be interesting to explore the idea whether the quantum seeds in such a context could have been relevant in the early universe. 

%Our remarks above, however, should not veil the fact that the question of a concrete physical relevance of the noninvertible phase in gravity theory remains very much open.
 
\begin{acknowledgments}
%\acknowledgments 
Support (in part) of the SERB, DST, Govt. of India., through the MATRICS project grant MTR/2021/000008 is gratefully acknowledged. I feel deeply indebted to Madhavan Varadarajan for very rewarding discussions on ref.\cite{peldan} in particular, and also to J. Fernando Barbero and Romesh Kaul for numerous useful conversations regarding degenerate metrics in general. I also thank the organizers and participants of the `10th Tux Workshop on Quantum Gravity', Austria, Feb 13-17, 2023, where part of this work was first presented.
\end{acknowledgments}

\end{document}